\documentclass[preprintnumbers, floatfix, onecolumn,
preprintnumbers, letterpaper, superscriptaddress,nofootinbib]{revtex4}
\usepackage{eurosym}
\usepackage{amsfonts}
\usepackage{amsmath}
\usepackage{amssymb,epsf}
\usepackage{latexsym}
\usepackage{graphicx,epsfig}
\usepackage{amssymb}
\usepackage{subfigure}
\usepackage[colorlinks=true,citecolor=blue,linkcolor=blue,urlcolor=black]{hyperref}
\graphicspath{{Images/}}

\begin{document}

\title{Optical properties of Born-Infeld-dilaton-Lifshitz holographic superconductors}
\author{M. Kord Zangeneh}
\email{mkzangeneh@scu.ac.ir}
\affiliation{Physics Department, Faculty of Science, Shahid Chamran University of Ahvaz,
Ahvaz 61357-43135, Iran}
\affiliation{Research Institute for Astronomy and Astrophysics of Maragha (RIAAM), P. O.
Box: 55134-441, Maragha, Iran}
\affiliation{Center for Research on Laser and Plasma, Shahid Chamran University of Ahvaz,
Ahvaz, Iran}
\author{S. S. Hashemi}
\affiliation{Physics Department, Shahid Beheshti University, Evin, Tehran 19839, Iran}
\affiliation{Center of Astronomy and Astrophysics, Department of Physics and Astronomy,
Shanghai Jiao Tong University, Shanghai 200240, China}
\author{A. Dehyadegari}
\email{adehyadegari@shirazu.ac.ir}
\affiliation{Physics Department and Biruni Observatory, Shiraz University, Shiraz 71454,
Iran}
\author{A. Sheykhi}
\email{asheykhi@shirazu.ac.ir}
\affiliation{Physics Department and Biruni Observatory, Shiraz University, Shiraz 71454,
Iran}
\affiliation{Research Institute for Astronomy and Astrophysics of Maragha (RIAAM), P. O.
Box: 55134-441, Maragha, Iran}
\author{B. Wang}
\email{wang\_b@sjtu.edu.cn}
\affiliation{Center of Astronomy and Astrophysics, Department of Physics and Astronomy,
Shanghai Jiao Tong University, Shanghai 200240, China}
\affiliation{Center for Gravitation and Cosmology, College of Physical Science and
Technology, Yangzhou University, Yangzhou 225009, China}

\begin{abstract}
In this paper, we first study the Lifshitz-dilaton holographic
superconductors with nonlinear Born-Infeld (BI) gauge field and obtain the
critical temperature of the system for different values of Lifshitz
dynamical exponent, $z$, and nonlinear parameter $b$. We find that for fixed
value of $b$, the critical temperature decreases with increasing $z$. This
indicates that the increase of anisotropy between space and time prevents
the phase transition. Also, for fixed value of $z$, the critical temperature
decrease with increasing $b$. Then, we investigate the optical properties of
($2+1$) and ($3+1$)-dimensional BI-Lifshitz holographic superconductors in
the the presence of dilaton field. We explore the refractive index of the
system. For $z=1$ and $(2+1)$-dimensional holographic superconductor, we
observe negative real part for permittivity \textrm{Re}$[\epsilon]$ as
frequency $\omega $ decreases. Thus, in low frequency region our
superconductor exhibit metamaterial property. This behavior is independent
of the nonlinear parameter and can be seen for either linear ($b=0$) and
nonlinear ($b\neq 0$) electrodynamics. Interestingly, for ($3+1$%
)-dimensional Lifshitz-dilaton holographic superconductors, we observe
metamaterial behavior neither in the presence of linear nor nonlinear
electrodynamics.
\end{abstract}

\maketitle

\section{Introduction}

The correspondence between gauge fields living on the boundary of a
spacetime and the gravity in the bulk, called \textit{gauge/gravity }
duality, provides a powerful tool for studying the strongly coupled systems
in quantum field theory \cite{Mal,Wit}. According to this dictionary, one
can effectively calculate correlation functions in a strongly interacting
field theory by using a dual classical gravity description. A new
application of this duality, called holographic superconductor, was recently
proposed in order to shed light on the understanding the mechanism governing
the high-temperature superconductors in condensed-matter physics \cite%
{Hartnol}. In \cite{Hartnol,Har2,Hor}, by including the Abelian Higgs model
within the AdS black hole spacetime, holographic superconductors have been
built. Decreasing the Hawking temperature to a critical value, the black
hole exhibits unstability against small perturbations and by condensing some
field, grows hair to make the system stable. This process can be adjudged as
the holographic description of the superconducting phase transition. The
(asymptotically) AdS black hole spacetime is taken as the starting point for
this kind of building for holographic superconductors. The studies on the
holographic superconductor have got a lot of attentions (see for example 
\cite{Mus,Gr, RGC1,P.BGRL, P.MRM, P.CW,
P.ZGJZ,RGC2,RGC3,RGC4,XHWang,Wang2,Wang3,Wang4,Wang5,Wang6,
Wang7,JJ1,JJ3,SG2,Shey2,Shey3,Shey4,SheyLN,SheyPM,1609.08402} and references
therein). When the gauge field is in the form of nonlinear BI
electrodynamics, the holographic superconductors have been explored in \cite%
{JJ2,Gan1,Gan2,Zh,1306.0064,Lai,SG1,Shey1,1211.0904,1710.09630}. It was
argued that the nonlinear parameter decrease the critical temperature of the
superconductor and make the condensation harder, while the critical exponent
associated with the condensation near the critical temperature has still the
universal value $1/2$ of the mean field theory \cite%
{JJ2,Gan1,Gan2,Zh,1306.0064,Lai,SG1,Shey1}.

On the other hand, since in condensed matter physics a dynamical exponent
appears near the critical point, the studies on Lifshitz spacetimes have
attained a lot of interests. It was shown that the near critical point
dynamics of such systems can be described by a relativistic conformal field
theory or a more subtle scaling theory respecting the Lifshitz symmetry $%
t\rightarrow \lambda ^{z}t,$ $\vec{\mathbf{x}}\rightarrow \lambda \vec{%
\mathbf{x}}$ \cite{Lif}. Various aspects of Lifshitz spacetimes have been
explored in the literature \cite%
{masssol,Deh1,hcc,toplif,peet,tario,Deh2,Deh3,hololif,vislif,
CondLif,KordLif1,KordLif2,KordLif3,KordLif4}. The holographic
superconductors were also extended to the context of Lifshitz black holes 
\cite{Bry,Sin}. It has been argued that despite quite different behaviors of
asymptotically AdS and Lifshitz geometries as a result of existing dynamical
exponent, the holographic Lifshitz superconductors basically behave the same
qualitatively as AdS ones \cite{Bu}. Effects of dynamical exponent $z$ on
the holographic Lifshitz superconductor models have been disclosed both
numerically and analytically in \cite{Lu}. Holographic superconducting phase
transitions of a system dual to Yang-Mills field coupled to an axion as
probes of black hole with arbitrary Lifshitz scaling have been studied with $%
p_{x}+ip_{y}$ condensate \cite{Tall}. The properties of the Weyl holographic
superconductor in the Lifshitz black hole background were explored in \cite%
{Mirza1}. It was shown that the critical temperature of the Weyl
superconductor decreases with increasing Lifshitz dynamical exponent, $z$,
indicating that condensation becomes difficult \cite{Mirza1}. Analytical
study on the properties of the $s$-wave holographic superconductor in the
presence of Exponential nonlinear electrodynamics in the Lifshitz black hole
background in four-dimensions has been done in \cite{Mirza2}.

It is also interesting to investigate the optical properties of holographic
superconductors\cite{1008.0720,1107.1242,1305.6273,1404.2190,1411.6405}. It
has been argued that no negative refractive index is seen for
Lifshitz-dilaton-Maxwell holographic ($2+1$)-dimensional superconductors in
probe limit \cite{1305.6273}. Here, we re-investigate this case and show
that one could find negative index of refraction albeit just for Lifshitz
exponent $z=1$. We also extend the study of \cite{1305.6273} in probe limit
to ($3+1$)-dimensional Lifshitz-dilaton superconductors as well as nonlinear
BI electrodynamics. We show that one could observe negative refractive index
for nonlinearly charged ($2+1$)-dimensional superconductors just with $z=1$
whereas no such effect is occurred for ($3+1$)-dimensional Lifshitz-dilaton
superconductors neither in the presence of linear nor nonlinear
electrodynamics for any $z$. The latter result is in agreement with the
conclusions of \cite{1008.0720} and \cite{1107.1242} in AdS case and further
reveal that neither Lifshitz scaling nor presence of nonlinear
electrodynamics help ($3+1$)-dimensional Lifshitz-dilaton holographic
superconductors to exhibit negative refraction phenomenon.

This paper is organized as follows. In section \ref{sec2}, we describe the
holographic superconductors in Lifshitz-dilaton setup and in the probe
limit, when the gauge field is in the form of BI nonlinear electrodynamics.
In section \ref{sec3}, we investigate the optical properties of
BI-dilaton-Lifshitz holographic superconductors. In particular, we shall
study the behavior of refractive index for our holographic superconductors.
The last section is devoted to summary and conclusions.

\section{Holographic setup of BI-dilaton-Lifshitz superconductors\label{sec2}%
}

In this section, we will present the holographic setup for dilaton-Lifshitz
superconductors in the presence of BI electrodynamics. The $D$-dimensional
metric which is invariant under the dynamical scaling $t\rightarrow \lambda
^{z}t$, $\vec{\mathbf{x}}\rightarrow \lambda \vec{\mathbf{x}}$ where $z$ is
dynamical critical exponent is given by \cite{Lif}%
\begin{equation}
\mathrm{d}s^{2}=L^{2}\left( -r^{2z}\mathrm{d}t^{2}+r^{2}\mathrm{d}\vec{%
\mathbf{x}}^{2}+\frac{\mathrm{d}r^{2}}{r^{2}}\right) ,  \label{1}
\end{equation}%
where $\mathrm{d}\vec{\mathbf{x}}^{2}=\sum_{i=1}^{d}\mathrm{d}x_{i}^{2}$, $%
D=d+2$ and $0<r<\infty $. The AdS $_{d+2}$ can be recovered by setting $z=1$
in metric (\ref{1}). (\ref{1}) is a solution of the action \cite{0905.2678} 
\begin{equation}
S=\frac{1}{16\pi G_{d+2}}\int \mathrm{d}^{d+2}\sqrt{-g}\left( R-2\Lambda -%
\frac{1}{2}\left( \partial \varphi \right) ^{2}-\frac{1}{4}e^{\lambda
\varphi }H_{\mu \nu }H^{\mu \nu }\right) ,  \label{action}
\end{equation}%
where $R$ is the Ricci scalar, $\Lambda $ is the cosmological constant, $%
\varphi $ is dilaton scalar field and $H_{\mu \nu }=\partial _{\lbrack \mu
}B_{\nu ]}$ where $B_{\nu }$ is a gauge potential, provided%
\begin{gather}
H_{rt}=\sqrt{2\left( z-1\right) \left( z+d\right) }Lr^{z+d-1},\text{ \ \ \ \ 
}e^{\lambda \varphi }=r^{-2d},  \notag \\
\quad \Lambda =-\frac{(z+d-1)(z+d)}{2L^{2}},\text{ \ \ \ \ }\lambda =\sqrt{%
\frac{2d}{z-1}}.
\end{gather}

The finite temperature generalization of metric (\ref{1}) which is an
asymptotic Lifshitz black hole is \cite{0905.2678} 
\begin{equation}
\mathrm{d}s^{2}=L^{2}\left( -r^{2z}f\left( r\right) \mathrm{d}t^{2}+\frac{1}{%
r^{2}f\left( r\right) }\mathrm{d}r^{2}+r^{2}\mathrm{d}\vec{\mathbf{x}}%
^{2}\right) ,  \label{met}
\end{equation}%
where 
\begin{equation}
f\left( r\right) =1-\frac{r_{+}^{z+d}}{r^{z+d}},
\end{equation}%
where $r_{+}$ is the black hole horizon. The Hawking temperature of this
black hole is%
\begin{equation}
T=\frac{(z+d)r_{+}^{z}}{4\pi }.
\end{equation}%
\begin{figure}[t]
\includegraphics[width=.4\textwidth]{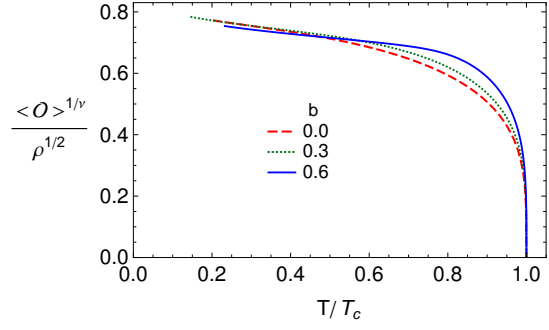}
\caption{The behavior of $\left\langle \mathcal{O}\right\rangle ^{1/\protect%
\nu }/\protect\rho ^{1/d}$ vs. $T/T_{c}$ for the case $d=2$ and $z=2$ where $%
\protect\nu =2$ with $b=0,$ $0.3$ and $0.6$.}
\label{fig1}
\end{figure}

\begin{figure}[t]
\includegraphics[width=.4\textwidth]{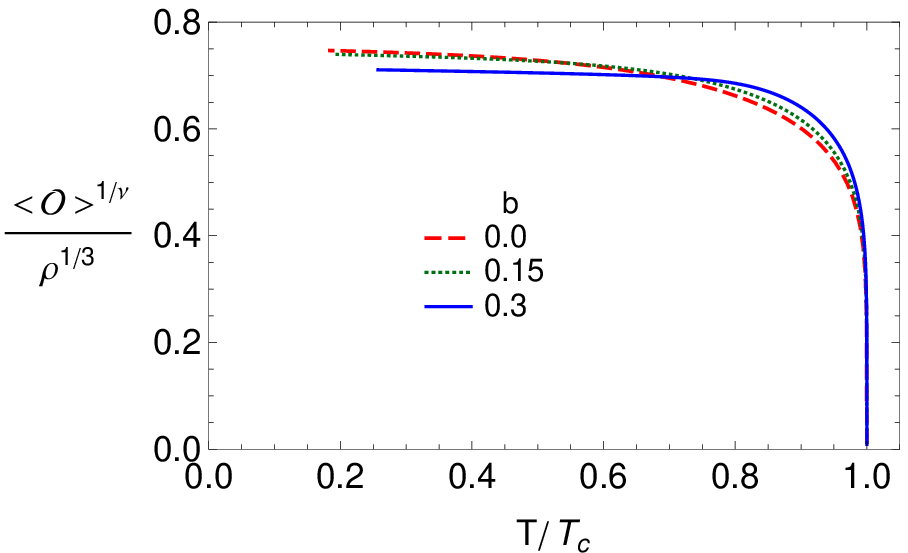}
\caption{The behavior of $\left\langle \mathcal{O}\right\rangle ^{1/\protect%
\nu }/\protect\rho ^{1/d}$ vs. $T/T_{c}$ for the case $d=3$ and $z=2$ where $%
\protect\nu =3$ with $b=0,$ $0.15$ and $0.3$.}
\label{fig2}
\end{figure}

\begin{figure}[t]
\includegraphics[width=.4\textwidth]{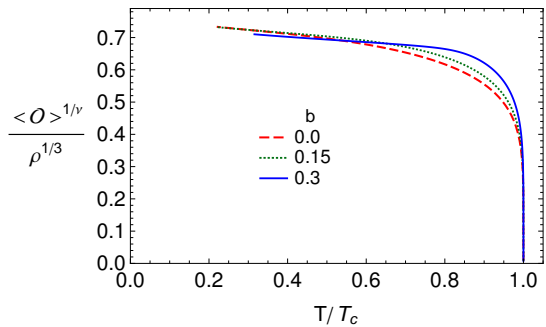}
\caption{The behavior of $\left\langle \mathcal{O}\right\rangle ^{1/\protect%
\nu }/\protect\rho ^{1/d}$ vs. $T/T_{c}$ for the case $d=3$ and $z=3$ where $%
\protect\nu =3$ with $b=0,$ $0.15$ and $0.3$.}
\label{fig3}
\end{figure}
Here, we intend to build a Lifshitz holographic superconductor in the
presence of BI electrodynamics in probe limit. Therefore, we consider the
following Lagrangian density for electrodynamics and scalar field%
\begin{equation}
\mathcal{L}_{m}=\mathcal{L}_{BI}-|\nabla _{\mu }\psi -iqA_{\mu }\psi
|^{2}-m^{2}\psi ^{2},  \label{Lm}
\end{equation}%
where $\psi $ is a charged complex scalar field, $A_{\mu }$ is gauge field
and $q$ and $m$ are the charge and the mass of the scalar field
respectively. $\mathcal{L}_{BI}$ is Lagrangian density of the BI
electrodynamics%
\begin{equation}
\mathcal{L}_{BI}=\frac{1}{b^{2}}\left( 1-\sqrt{1+\frac{b^{2}F_{\mu \nu
}F^{\mu \nu }}{2}}\right) ,  \label{LBI}
\end{equation}%
where $F_{\mu \nu }=\partial _{\lbrack \mu }A_{\nu ]}$ is the electrodynamic
tensor and $b$ is the BI nonlinear parameter.In the limit $b\rightarrow 0$, (%
\ref{LBI}) reduces to the linear Maxwell case. We choose the ansatz \cite%
{Hartnol} 
\begin{equation}
\psi =\psi \left( r\right) ,\quad A_{\mu }\mathrm{d}x^{\mu }=\phi \left(
r\right) \mathrm{d}t.
\end{equation}%
Varying (\ref{Lm}), the equations of motion for the complex scalar field $%
\psi \left( r\right) $ and the electric potential $\phi (r)$ in the
background (\ref{met}) are%
\begin{equation}
\psi ^{\prime \prime }+\left( \frac{d+z+1}{r}+\frac{f^{\prime }}{f}\right)
\psi ^{\prime }+\frac{q^{2}\phi ^{2}}{r^{2z+2}f^{2}}\psi -\frac{m^{2}L^{2}}{%
r^{2}f}\psi =0,  \label{9}
\end{equation}%
and 
\begin{eqnarray}
\phi ^{\prime \prime }+\frac{d-z+1}{r}\phi ^{\prime }-\frac{2q^{2}L^{2}\psi
^{2}\phi }{r^{2}f}\sqrt{1-\frac{b^{2}\phi ^{\prime 2}}{L^{4}r^{2z-2}}}\left(
1-\frac{b^{2}}{L^{4}r^{2(z-1)}}\phi ^{\prime 2}\right) -\frac{b^{2}d}{%
L^{4}r^{2z-1}}\phi ^{\prime 3} &=&0,  \notag \\
&&
\end{eqnarray}%
where prime stands for derivative with respect to $r$. It is difficult to
find the analytical solutions to these nonlinear equations. However, they
are reachable numerically. In order to solve the above two equations
numerically, we specify the boundary conditions for $\psi $ and $\phi $ near
the horizon of black hole $r=r_{+}$ and in the spatial infinity $%
r\rightarrow \infty $. In order to satisfy the regularity condition at the
horizon we demand $\phi (r_{+})=0$. Also, $\psi (r_{+})$ should be regular
at horizon. At the boundary $r\rightarrow \infty $, the complex scalar field 
$\psi $ and the scalar potential $\phi $ behave as%
\begin{equation}
\psi (r)=\left\{ 
\begin{tabular}{ll}
$\left( \psi ^{(2)}+\psi ^{(1)}\ln \xi r\right) r^{-\nu }+\cdots $, & for $%
\nu =\nu _{+}=\nu _{-}$, \\ 
&  \\ 
$\psi ^{(1)}r^{-\nu _{-}}+\psi ^{(2)}r^{-\nu _{+}}+\cdots $, & other cases,%
\end{tabular}%
\right.  \label{11}
\end{equation}%
where%
\begin{equation*}
\nu _{\pm }=\frac{z+d\pm \sqrt{(z+d)^{2}+4L^{2}m^{2}}}{2},
\end{equation*}%
and 
\begin{equation}
\phi (r)=\left\{ 
\begin{tabular}{ll}
$\mu -\rho r^{z-d}+\cdots $, & for $z<d$, \\ 
&  \\ 
$\mu -\rho \ln \xi r+\cdots $, & for $z=d$,%
\end{tabular}%
\right.  \label{12}
\end{equation}%
where $\psi ^{(1)}$, $\psi ^{(2)}$, $\mu $, $\rho $ and $\xi $ are
constants. According to gauge/gravity duality dictionary, $\psi ^{(1)}$ and $%
\psi ^{(2)}$ correspond to the vacuum expectation value of the dual operator 
$\mathcal{O}$ dual to the scalar field and $\mu $ and $\rho $ are chemical
potential and charge density of dual field theory respectively. 
\begin{table}[t]
\caption{The values of critical temperature for different $z$ and $b$.}
\label{tab1}
\begin{center}
\begin{tabular}{|c|c|c|c|c|c|c|c|c|c|c|c|c|c|}
\cline{1-4}\cline{6-9}\cline{11-14}
\multicolumn{4}{|l|}{$b=0$} &  & \multicolumn{4}{|l|}{$b=0.3$} &  & 
\multicolumn{4}{|l|}{$b=0.6$} \\ \cline{1-4}\cline{6-9}\cline{11-14}
$d$ & $z$ & $\nu $ & $T_{c}/\rho ^{z/d}$ &  & $d$ & $z$ & $\nu $ & $%
T_{c}/\rho ^{z/d}$ &  & $d$ & $z$ & $\nu $ & $T_{c}/\rho ^{z/d}$ \\ 
\cline{1-4}\cline{6-9}\cline{11-14}
$2$ & $1$ & $2$ & $0.1184$ &  & $2$ & $1$ & $2$ & $0.1023$ &  & $2$ & $1$ & $%
2$ & $0.0662$ \\ \cline{1-4}\cline{6-9}\cline{11-14}
$2$ & $2$ & $2$ & $0.0678$ &  & $2$ & $2$ & $2$ & $0.0475$ &  & $2$ & $2$ & $%
2$ & $0.0188$ \\ \cline{1-4}\cline{6-9}\cline{11-14}
$3$ & $1$ & $2$ & $0.2526$ &  & $3$ & $1$ & $2$ & $0.2336$ &  & $3$ & $1$ & $%
2$ & $0.1914$ \\ \cline{1-4}\cline{6-9}\cline{11-14}
$3$ & $1$ & $3$ & $0.1980$ &  & $3$ & $1$ & $3$ & $0.1331$ &  & $3$ & $1$ & $%
3$ & $0.0395$ \\ \cline{1-4}\cline{6-9}\cline{11-14}
$3$ & $2$ & $3$ & $0.0871$ &  & $3$ & $2$ & $3$ & $0.0295$ &  & $3$ & $2$ & $%
3$ & $0.0008$ \\ \cline{1-4}\cline{6-9}\cline{11-14}
$3$ & $3$ & $3$ & $0.0452$ &  & $3$ & $3$ & $3$ & $0.0094$ &  & $3$ & $3$ & $%
3$ & $0.0004$ \\ \cline{1-4}\cline{6-9}\cline{11-14}
\end{tabular}%
\end{center}
\end{table}
We impose $\psi ^{(1)}$ equals zero in order to make the superconducting
phase transition a spontaneous breaking of symmetry \cite{Lu}. Explicitly
the modified BF bound for the scalar mass is $m^{2}\geqslant -(z+d)^{2}/4$.
Throughout this paper when we discuss the black hole backgrounds we assume
that $\rho $ is fixed. Also $\nu $ denotes $\nu _{+}$. For our numerical
calculations we focus on the cases $z=2$ in $d=2$ and $z=2,3$ in $d=3$ for
different values of the parameter $b$. In our numerical calculations, we fix
the dimension $\nu $, in order to see the influence of the dynamical
critical exponent $z$ and nonlinear parameter $b$. Figs. \ref{fig1}, \ref%
{fig2} and \ref{fig3} are depicted for the condensation as a function of
temperature for various $z$ and $b$. It is seen from the figures that the
condensation decreases as the parameter $z$ increases. In table \ref{tab1},
we show the values of $T_{c}/\rho ^{z/d}$ for the cases $z=2$ and $z=3$ with
fixed $\nu $'s. As it is seen from the table \ref{tab1}, for each case as we
increase $z$, the critical temperature decreases, showing that the increase
of anisotropy between space and time prevents the phase transition. This can
be explained as follows. As it is seen from equation (\ref{9}) as the
dynamical critical exponent $z$ increases, the effective mass of the scalar
field increases near the horizon. This causes a lower critical temperature
as $z$ is increased \cite{Lu}.

\section{Optical properties of BI-Dilaton-Lifshitz holographic
superconductors \label{sec3}}

Here, we will investigate the optical properties of BI-dilaton-Lifshitz
holographic superconductors. Specially, we will study the behavior of
refractive index (focusing on the negativity of it) for our holographic
superconductors. In order to do this, we will study the behavior of
Depine-Lakhtakia (DL) index \cite{DL}

\begin{equation}
n_{DL}=\left\vert \epsilon \right\vert \mathrm{Re}[\mu ]+\left\vert \mu
\right\vert \mathrm{Re}[\epsilon ],  \label{nDL}
\end{equation}%
where $\epsilon $ is the permittivity and $\mu $ is effective permeability.
The negativity of DL index shows that the phase velocity and energy flow are
in opposite directions in the medium. It implies that the refractive index
of the system is negative. It is clear from Eq. (\ref{nDL}) that in order to
compute DL index we need to obtain $\epsilon $ and $\mu $. To determine $%
\epsilon $ and $\mu $ corresponding to the superconductor, we apply an
external electric field by turning on $A_{j}=A_{j}\left( r\right)
e^{-i\omega t+iK_{k}x_{k}}$ where $j$ and $k$ stand for two directions which
are perpendicular to each other e.g. $x$ and $y$, $\omega $ is the frequency
and $K_{k}$ is the momentum. From now on, we replace $A_{j}$ with $A$ and $%
K_{k}$ with $K$ for abbreviation. Since we are in probe limit, the equation
of motion for $A$ is decoupled from other ones and reads%
\begin{equation}
A^{\prime \prime }+\left( \frac{d+z-1}{r}+\frac{f^{\prime }}{f}-\frac{db^{2}%
}{L^{4}r^{2z-1}}\phi ^{\prime 2}\right) A^{\prime }+\left( \frac{\omega ^{2}%
}{r^{2(z+1)}f^{2}}-\frac{K^{2}}{r^{4}f}\right) A-\frac{2L^{2}q^{2}\psi ^{2}}{%
r^{2}f}\sqrt{1-\frac{b^{2}\phi ^{\prime 2}}{L^{4}r^{2z-2}}}\left( A-\frac{%
b^{2}\phi \phi ^{\prime }}{L^{4}r^{2(z-1)}}A^{\prime }\right) =0.
\label{AFE}
\end{equation}%
We restrict our study here to $d=2$ and $3$ dimensions. The behavior of
\bigskip $A$ as $r$ goes to infinity is then%
\begin{equation}
\lim_{r\rightarrow \infty }A=\left\{ 
\begin{array}{ll}
A_{0}+A_{1}r^{-z}+\cdots , & d=2, \\ 
&  \\ 
A_{0}+A_{1}r^{-z-1}+\cdots , & d=3\text{ }(z\neq 1), \\ 
&  \\ 
A_{0}\left( 1+\frac{1}{2}\omega ^{2}r^{-2}\ln \xi r\right)
+A_{1}r^{-2}+\cdots , & d=3\text{ }(z=1).%
\end{array}%
\right.  \label{infA}
\end{equation}%
Also, $A$ behaves as $f^{-i\omega /4\pi T}$ near the horizon (by employing
the ingoing wave boundary condition near the horizon). According to
gauge/gravity duality dictionary, one could holographically interpret $A_{0} 
$ and $A_{1}$ as dual source and the expectation value of boundary current.
Therefore, the current-current ($J_{j}-J_{j}$) correlator $G_{jj}$ (the
response to $A$) has a relation $G_{jj}=A_{1}/A_{0}$ \cite{horo}. On the
other hand, $\epsilon $ and $\mu $ has relation with $G_{jj}$ so that \cite%
{landau}%
\begin{equation}
\epsilon \left( \omega \right) =1+4\pi \omega ^{-2}C^{2}G_{jj}^{0}\left(
\omega \right) ,  \label{eps}
\end{equation}%
\begin{equation}
\mu \left( \omega \right) =\left[ 1-4\pi C^{2}G_{jj}^{2}\left( \omega
\right) \right] ^{-1},  \label{mu}
\end{equation}%
where $C$ is the electromagnetic coupling constant (which we set it to unity
in our calculations) and $G_{jj}^{0}\left( \omega \right) $ and $%
G_{jj}^{2}\left( \omega \right) $ come from casting the $K$-dependent $%
G_{jj} $ \cite{son}%
\begin{equation}
G_{jj}\left( \omega ,K\right) =G_{jj}^{0}\left( \omega \right)
+K^{2}G_{jj}^{2}\left( \omega \right) +\cdots .  \label{Gexpand}
\end{equation}%
To find the relation between $G_{jj}^{0}$ and $G_{jj}^{2}$ with $A_{0}$ and $%
A_{1}$, we expand $A$ in powers of $K$ in the same way as $G_{jj}$:

\begin{equation}
A\left( r\right) =A^{0}\left( r\right) +K^{2}A^{2}\left( r\right) +\cdots .
\end{equation}%
Then, the corresponding field equations come from Eq. (\ref{AFE}) are

\begin{equation}
A^{0\prime \prime }+\left( \frac{d+z-1}{r}+\frac{f^{\prime }}{f}-\frac{db^{2}%
}{L^{4}r^{2z-1}}\phi ^{\prime 2}\right) A^{0\prime }+\frac{\omega ^{2}}{%
r^{2(z+1)}f^{2}}A^{0}-\frac{2L^{2}q^{2}\psi ^{2}}{r^{2}f}\sqrt{1-\frac{%
b^{2}\phi ^{\prime 2}}{L^{4}r^{2z-2}}}\left( A^{0}-\frac{b^{2}\phi \phi
^{\prime }}{L^{4}r^{2(z-1)}}A^{0\prime }\right) =0,  \label{FEA0}
\end{equation}%
and

\begin{equation}
A^{2\prime \prime }+\left( \frac{d+z-1}{r}+\frac{f^{\prime }}{f}-\frac{db^{2}%
}{L^{4}r^{2z-1}}\phi ^{\prime 2}\right) A^{2\prime }+\frac{\omega ^{2}}{%
r^{2(z+1)}f^{2}}A^{2}-\frac{2L^{2}q^{2}\psi ^{2}}{r^{2}f}\sqrt{1-\frac{%
b^{2}\phi ^{\prime 2}}{L^{4}r^{2z-2}}}\left( A^{2}-\frac{b^{2}\phi \phi
^{\prime }}{L^{4}r^{2(z-1)}}A^{2\prime }\right) -\frac{A^{0}}{r^{4}f}=0.
\label{FEA2}
\end{equation}%
The behaviors of $A^{0}$ and $A^{2}$ as $r\rightarrow \infty $ are the same
as $A$ in Eq. (\ref{infA}). Finally, we have%
\begin{equation}
G_{jj}^{0}=\left\{ 
\begin{array}{ll}
\frac{zA_{1}^{0}}{A_{0}^{0}}, & d=2, \\ 
&  \\ 
\frac{(z+1)A_{1}^{0}}{A_{0}^{0}}, & d=3\text{ }(z\neq 1), \\ 
&  \\ 
\frac{A_{1}^{0}}{A_{0}^{0}}+c\omega ^{2}, & d=3\text{ }(z=1),%
\end{array}%
\right.
\end{equation}%
and%
\begin{equation}
G_{jj}^{2}=\left\{ 
\begin{array}{ll}
\frac{zA_{1}^{0}}{A_{0}^{0}}\left( \frac{A_{1}^{2}}{A_{1}^{0}}-\frac{%
A_{0}^{2}}{A_{0}^{0}}\right) , & d=2, \\ 
&  \\ 
\frac{(z+1)A_{1}^{0}}{A_{0}^{0}}\left( \frac{A_{1}^{2}}{A_{1}^{0}}-\frac{%
A_{0}^{2}}{A_{0}^{0}}\right) , & d=3\text{ }(z\neq 1), \\ 
&  \\ 
\frac{A_{1}^{0}}{A_{0}^{0}}\left( \frac{A_{1}^{2}}{A_{1}^{0}}-\frac{A_{0}^{2}%
}{A_{0}^{0}}\right) -c, & d=3\text{ }(z=1),%
\end{array}%
\right.
\end{equation}%
where $c$ could be fixed so that $G_{jj}^{0}$ vanishes at large frequencies
where the variation of external field is too rapid to be responded by the
system. Now, we can solve Eqs. (\ref{FEA0}) and (\ref{FEA2}) numerically to
find $G_{jj}^{0}$ and $G_{jj}^{2}$. Then, we compute the permittivity $%
\epsilon $ and effective permeability $\mu $. Eventually, using $\epsilon $
and $\mu $, we can calculate DL index $n_{DL}$ via Eq. (\ref{nDL}).

\subsection{Numerical results}

\begin{figure*}[t]
\centering{%
\subfigure[]{
   \label{fig4a}\includegraphics[width=.4\textwidth]{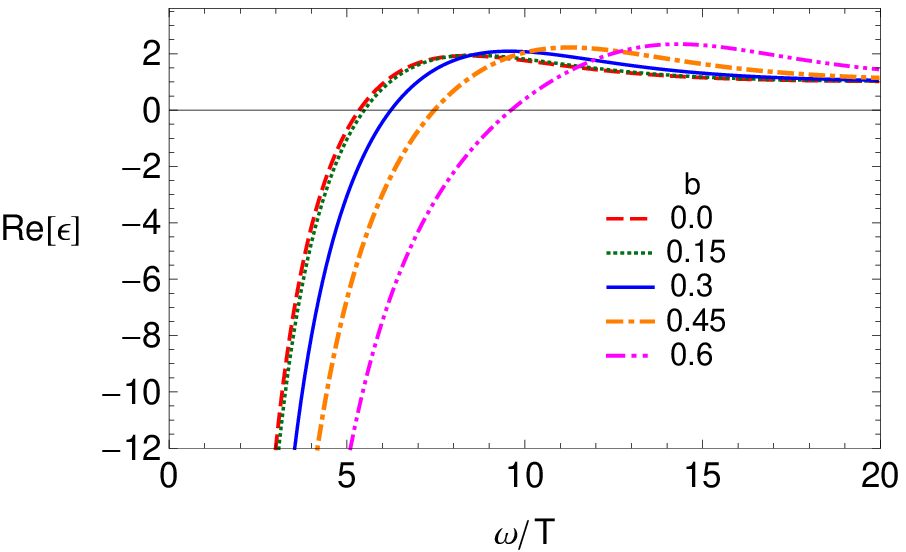}}} 
\subfigure[]{
   \label{fig4b}\includegraphics[width=.4\textwidth]{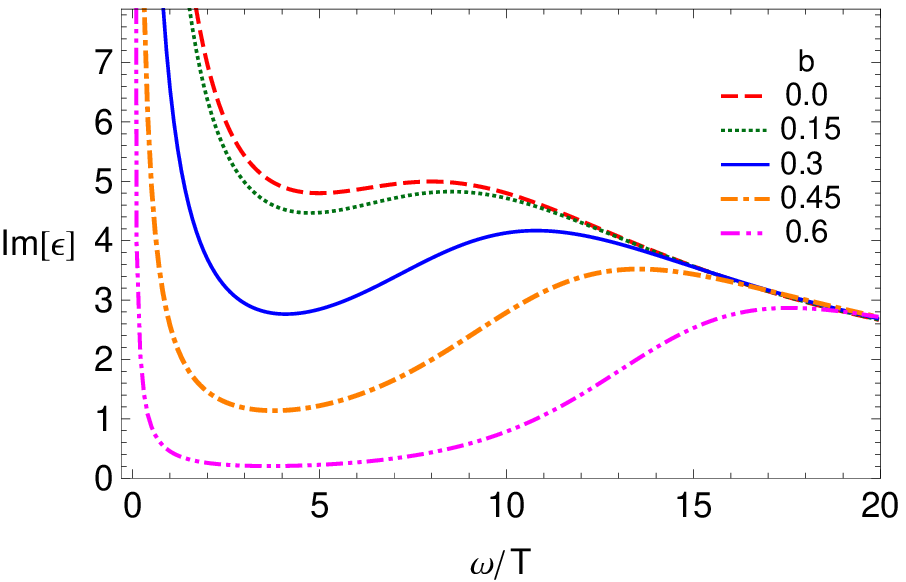}}
\caption{The behaviors of $\protect\epsilon $ as a function of $\protect%
\omega /T$ for $d=2$, $z=1$ at $T=0.7Tc$.}
\label{fig4}
\end{figure*}

\begin{figure*}[t]
\centering{%
\subfigure[]{
   \label{fig5a}\includegraphics[width=.4\textwidth]{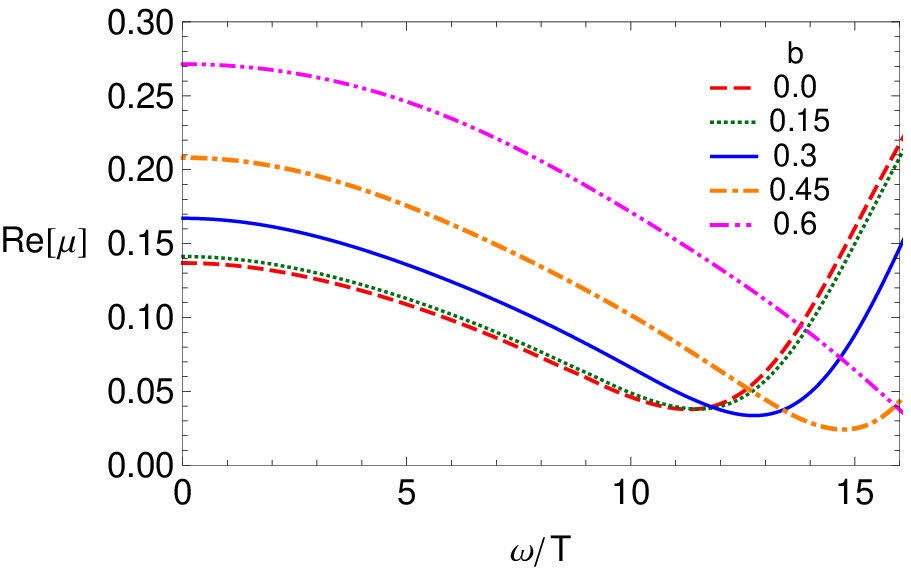}}} 
\subfigure[]{
   \label{fig5b}\includegraphics[width=.4\textwidth]{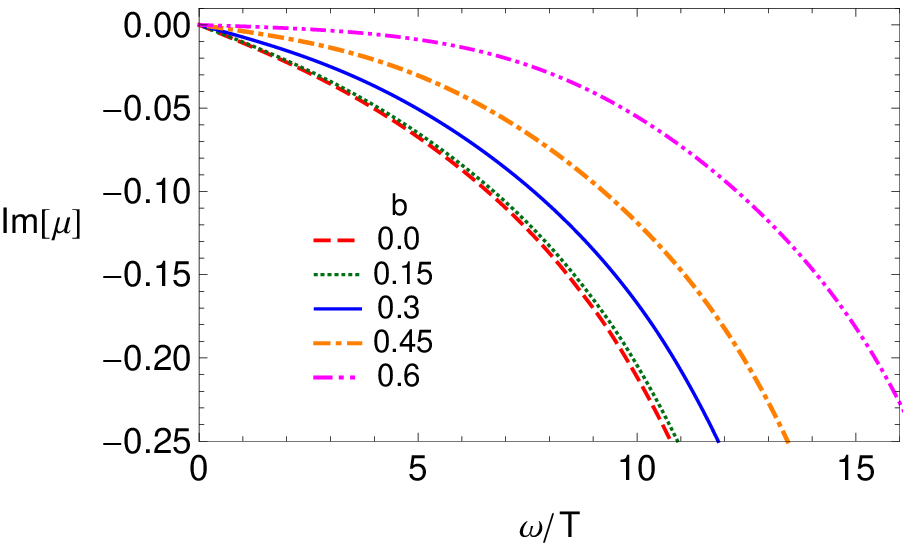}}
\caption{The behaviors of $\protect\mu $ as a function of $\protect\omega /T$
for $d=2$, $z=1$ at $T=0.7Tc$.}
\label{fig5}
\end{figure*}

\begin{figure*}[t]
\centering{%
\subfigure[]{
   \label{fig6a}\includegraphics[width=.4\textwidth]{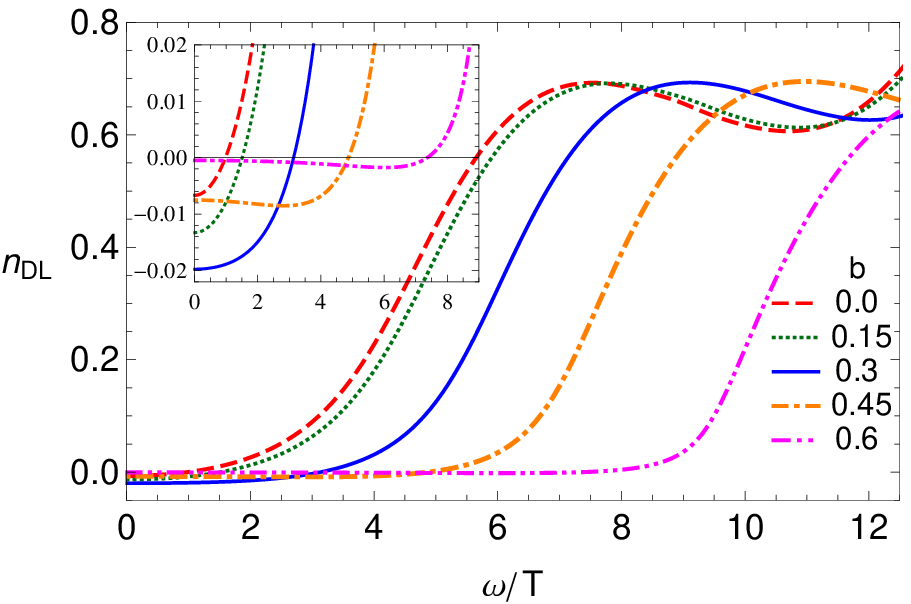}}} 
\subfigure[]{
   \label{fig6b}\includegraphics[width=.4\textwidth]{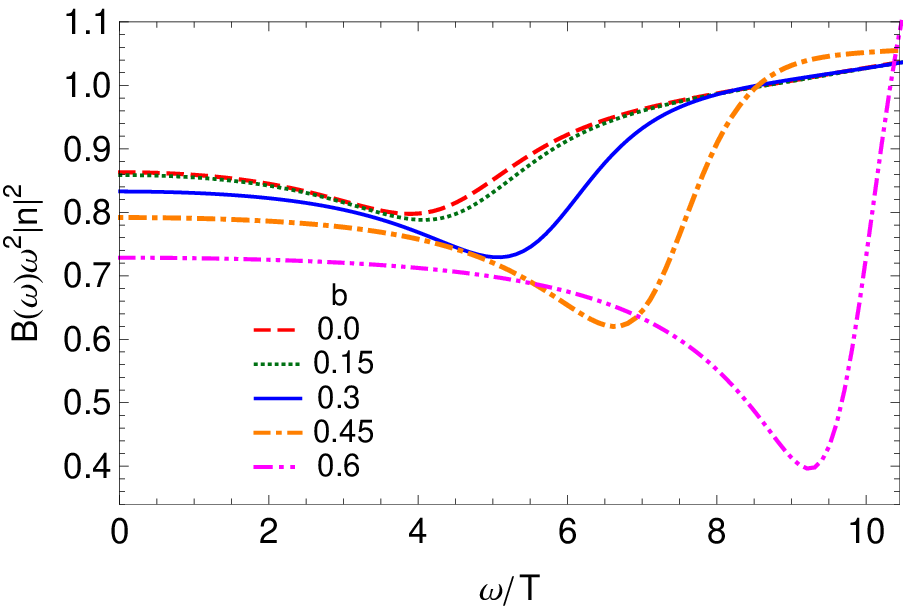}}
\caption{The behaviors of $n_{DL}$ and $B(\protect\omega )\protect\omega %
^{2}|n|^{2}$ as a function of $\protect\omega /T$ for $d=2$, $z=1$ at $%
T=0.7Tc$.}
\label{fig6}
\end{figure*}

\begin{figure*}[t]
\centering{%
\subfigure[]{
   \label{fig7a}\includegraphics[width=.4\textwidth]{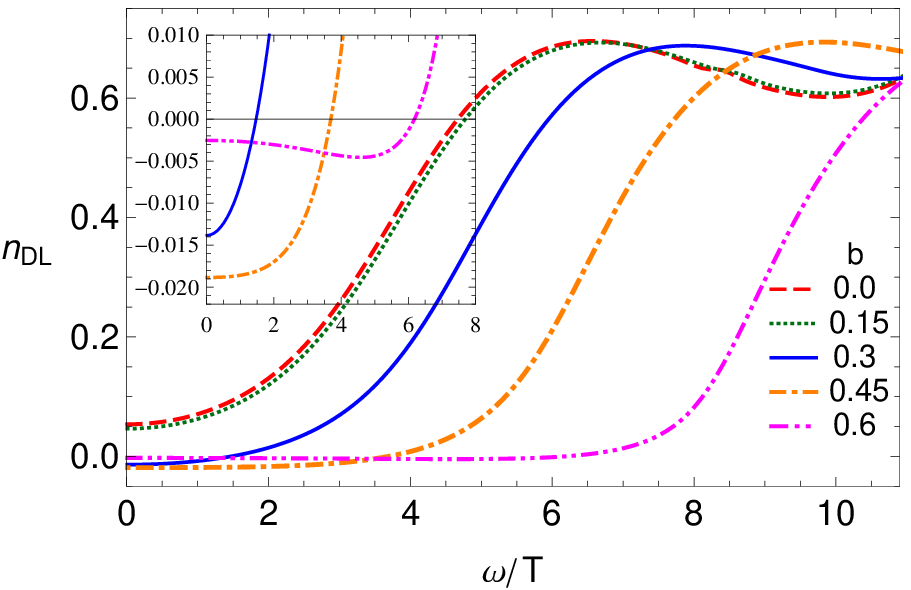}}} 
\subfigure[]{
   \label{fig7b}\includegraphics[width=.4\textwidth]{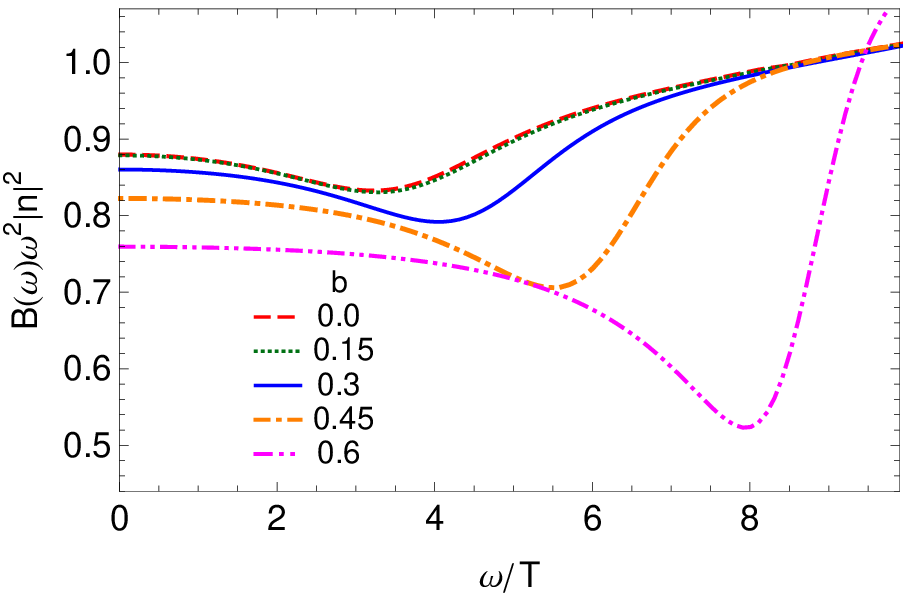}}
\caption{The behaviors of $n_{DL}$ and $B(\protect\omega )\protect\omega %
^{2}|n|^{2}$ as a function of $\protect\omega /T$ for $d=2$, $z=1$ at $%
T=0.78Tc$.}
\label{fig7}
\end{figure*}

\begin{figure}[t]
\includegraphics[width=.4\textwidth]{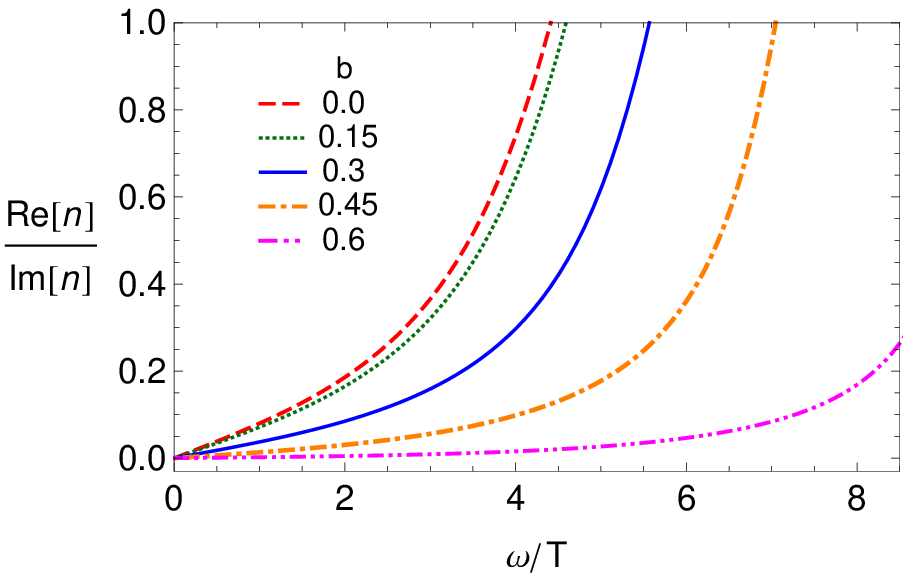}
\caption{The behaviors of $\mathrm{Re}(n)/\mathrm{Im}(n)$ as a function of $%
\protect\omega /T$ for $d=2$, $z=1$ at $T=0.7Tc$.}
\label{fig8}
\end{figure}

\begin{figure*}[t]
\centering{%
\subfigure[]{
   \label{fig9a}\includegraphics[width=.4\textwidth]{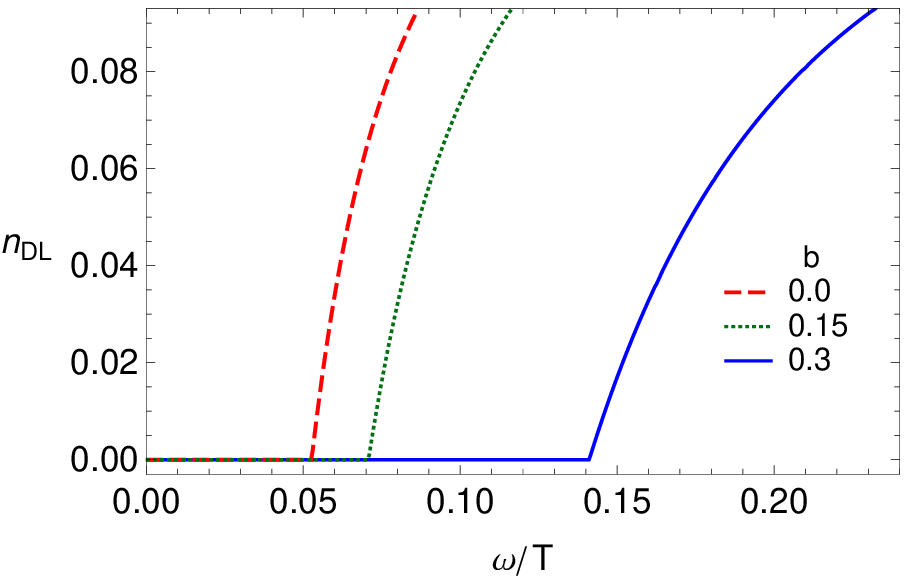}}} 
\subfigure[]{
   \label{fig9b}\includegraphics[width=.4\textwidth]{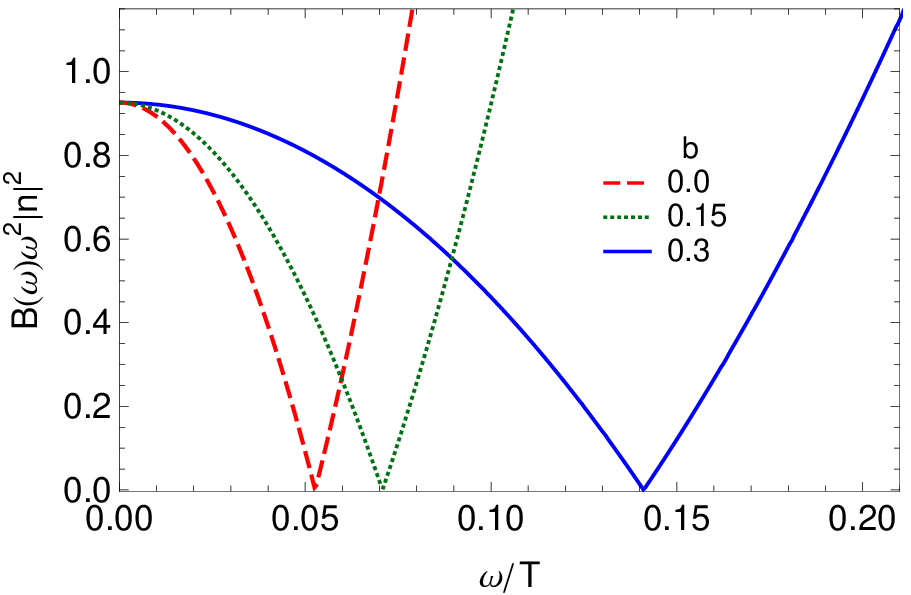}}
\caption{The behaviors of $n_{DL}$ and $B(\protect\omega )\protect\omega %
^{2}|n|^{2}$ as a function of $\protect\omega /T$ for $d=3$, $z=2$ at $%
T=0.7Tc$.}
\label{fig9}
\end{figure*}

Here, we present the optical properties of ($2+1$)- and ($3+1$)-dimensional
Lifshitz-dilaton holographic superconductors numerically. We start with
study the behavior of permittivity $\epsilon $ and effective permeability $%
\mu $ for ($2+1$)-dimensional superconductors with respect to frequency $%
\omega $. To ensure we are in superconducting phase, we fix the temperature $%
T$ to $0.7T_{c}$. As Fig. \ref{fig4a} shows, for $z=1$, we observe negative
real part for permittivity \textrm{Re}$[\epsilon ]$ as $\omega $ decreases.
Therefore, in low frequency region our superconductor exhibit metamaterial
property. This fact is held both in the presence of linear ($b=0$) and
nonlinear ($b\neq 0$) electrodynamics. As it is clear from Fig. \ref{fig4a},
the frequency at which $\epsilon $ changes the sign is larger for larger
values of $b$. The imaginary part of permittivity \textrm{Im}$[\epsilon ]$
is positive for all frequencies and has a pole as $\omega \rightarrow 0$
(see Fig. \ref{fig4b}). The behavior of effective permeability $\mu $ in
terms of $\omega $ is depicted in Fig. \ref{fig5}. The real part of it 
\textrm{Re}$[\mu ]$ is always positive as Fig. \ref{fig5a} shows. However,
the imaginary part of permeability \textrm{Im}$[\mu ]$ exhibits a seemingly
odd behavior. As one can see in Fig. \ref{fig5b}, it is negative for all
frequencies. Actually, this happens for all available probe limit analyses
in literature. This behavior may be problematic, although $\mu $ is not an
observable \cite{1107.1242,1404.2190}. However, the issue of sign of \textrm{%
Im}$[\mu ]$ is not yet a settled issue in literature \cite{1404.2190,mark}.
Including backreaction could make \textrm{Im}$[\mu ]$ positive \cite%
{1411.6405}.

Now, we turn to study the behavior of DL index $n_{DL}$ in terms of
frequency. This behavior is shown in Figs. \ref{fig6a} and \ref{fig7a} for $%
T=0.7T_{c}$ and $0.78T_{c}$, respectively. Before going on further, it is
necessary to give a comment on the validity of our analysis. As we discussed
in previous subsection, to calculate $n_{DL}$, we use $G_{jj}^{0}$ and $%
G_{jj}^{2}$ which are the coefficients of expansion of current-current
correlator $G_{jj}$ in terms of momentum $K$. Therefore, in order for the
series (\ref{Gexpand}) to be restricted to the first two terms, the
condition $\left\vert K^{2}G_{jj}^{2}/G_{jj}^{0}\right\vert \ll 1$ has to be
satisfied. By virtue of the relation $n^{2}=K^{2}/\omega ^{2}=\epsilon \mu $%
, one could rewrite the latter condition as $B\omega ^{2}\left\vert
n\right\vert ^{2}\ll 1$ where $B=\left\vert G_{jj}^{2}/G_{jj}^{0}\right\vert 
$. We depict $B\omega ^{2}\left\vert n\right\vert ^{2}$ in terms of
frequency in Figs. \ref{fig6b} and \ref{fig7b}. Figs. \ref{fig6a} and \ref%
{fig7a}, in agreement with Fig. \ref{fig4}, show that in low frequency
regime our superconductor exhibits metamaterial property namely, $n_{DL}<0$%
\footnote{%
Since in our numerical studies \textrm{Im}$[\mu ]<0$ (as all available probe
limit studies), there are some possibilities for interpreting $n_{DL}<0$
case \cite{1404.2190}. This issue is not settled \cite%
{1107.1242,1404.2190,mark} and one could interpret this case as metamaterial.}
. In low frequency regime, our analysis is valid (although not strictly
\footnote{%
This problem occurs in probe limit like the negativity of imaginary part of
permeability $\mu $ discussed later. It might be overcome by including
backreaction effect. It is an interesting and important issue for future
studies.}) as Figs. \ref{fig6b} and \ref{fig7b} confirm. As the value of
nonlinear parameter $b$ increases, the frequency at which the system goes
from positive refraction phase to negative refraction phase increases too.
This fact is again in agreement with the result of Fig. \ref{fig4}.
Comparing Figs. \ref{fig6a} and \ref{fig7a}, one could find out that by
increasing the temperature in superconducting phase, the negative $n_{DL}$
tends more and more to zero and positive values\footnote{%
It could be interesting to mention that a qualitatively similar result has
been reported in \cite{zhao}, albeit not for a superconductor.} and finally
the metamaterial behavior disappears. This fact occurs for the cases with
smaller $b$ sooner than the larger $b$ cases. Also, for a fixed $b$, the
frequency at which $n_{DL}$ changes the sign decreases as temperature
increases (see Figs. \ref{fig6a} and \ref{fig7a}).

It is also interesting to investigate the dissipation effects. These effects
could be obtained form the ratio $\mathrm{Re}[n]/\mathrm{Im}[n]$ which shows
the propagation to dissipation ratio. The behavior of $\mathrm{Re}[n]/%
\mathrm{Im}[n]$ in terms of frequency $\omega $ is depicted in Fig. \ref%
{fig8}. In a part of negative refraction domain, actually near the zero
frequency, this ratio tends to zero meaning that the electromagnetic wave is
virtually not propagated. This is not a desirable result, however, we would
keep in mind that we are in probe approximation. This problem as others
mentioned before may be the result of probe limit and like goes away by
including the effects of backreaction \cite{1107.1242}. In addition, unlike
the real metamaterials the signs of real and imaginary parts of $n$ are the
same. This also could likely be overcome if one works in backreacted regime.

To close this section, we point out here that for ($2+1$)-dimensional
Lifshitz-dilaton holographic superconductors we observed the metamaterial
behavior just for $z=1$ (the case we discussed above) both for linear and
nonlinear electrodynamics. Also, for ($3+1$)-dimensional Lifshitz-dilaton
holographic superconductors, we observed metamaterial behavior neither in
the presence of linear nor nonlinear electrodynamics. Fig. \ref{fig9} shows
the behavior of $n_{DL}$ in terms of $\omega $ in superconducting phase for $%
d=3$ and $z=2$ as an example. As one could see, $n_{DL}$ is always positive
for this case.

\section{Summary and conclusions}

The studies on the holographic superconductors theory have got a lot of
attentions in the past years. This new prescription of superconductor is
based on the \textit{gauge/gravity } duality which provides a powerful tool
for investigation the strongly coupled systems in quantum field theory. The
motivation for such studying comes from the unsolved problem in
condensed-matter physics. Indeed, it was argued that holographic
superconductor may shed light on the problem of high temperature
superconductors in condensed-matter physics \cite{Hartnol,Har2,Hor}.

In this paper, we have further continued the studies on the holographic
superconductors, by considering the framework of Lifshitz-dilaton black
holes when the gauge field is in the form of nonlinear BI electrodynamics.
We have considered the probe limit in which the gauge and scalar fields do
not affect on the background geometry. Due to the complexity of the field
equations, we determine the condensation value as well as the critical
temperature by using the numerical calculations. In numerical calculations,
we studied the cases with the dynamical critical exponent $z=2$ in $d=2$
(where $d+2$ is the spacetime dimensions) and $z=2,3$ with $d=3$ for
different values of the nonlinear parameter $b$. We plotted the condensation
as a function of the temperature for various $z$ and $b$. We find out that
with increasing $z$, the condensation operator decreases. We also
investigated the effects of $z$ and $b$ on the behavior of the critical
temperature. We observed that for fixed values of $b$, the critical
temperature decreases with increasing $z$, showing that the increase of
anisotropy between space and time prevents the phase transition. This is due
to the fact that as the dynamical critical exponent $z$ increases, the
effective mass of the scalar field increases near the horizon. Also, for
fixed value of $z$, the critical temperature decrease with increasing $b$.

We also numerically investigated the optical properties of ($2+1$) and ($3+1$%
)-dimensional Lifshitz-dilaton holographic superconductors. We observed that
in case of ($2+1)$-dimensions and $z=1$, the real part of permittivity 
\textrm{Re}$[\epsilon ]$ is negative as $\omega $ decreases. This implies
that in low frequency region, the superconductor exhibit metamaterial
property. This behavior is independent of the nonlinear parameter $b$ and
can be seen for both linear case ($b=0$) and nonlinear case ($b\neq 0$). We
find that the imaginary part of permittivity \textrm{Im}$[\epsilon ]$ is
positive for all frequencies and has a pole as $\omega \rightarrow 0$.
Besides, for ($3+1$)-dimensional Lifshitz-dilaton holographic
superconductors, we observed metamaterial behavior neither in the presence
of linear nor nonlinear electrodynamics.

\begin{acknowledgments}
MKZ would like to thank Shahid Chamran University of Ahvaz for supporting
this work. The work of MKZ has also been supported financially by Research
Institute for Astronomy \& Astrophysics of Maragha (RIAAM) under research
project No. 1/5237-49. AD and AS thank the research council of Shiraz
University.
\end{acknowledgments}


\begin{thebibliography}{99}
\bibitem{Mal} J. M. Maldacena, \textit{The large-N limit of superconformal
field theories and supergravity}, Adv. Theor. Math. Phys. \textbf{2} (1998)
231 [Int. J. Theor. Phys. \textbf{38} (1999) 1113] [hep-th/9711200].

\bibitem{Wit} E. Witten, \textit{Anti de Sitter space and holography}, Adv.
Theor. Math. Phys. \textbf{2} (1998) 253 [hep-th/9802150].

\bibitem{Hartnol} S. A. Hartnoll, C. P. Herzog and G. T. Horowitz, \textit{%
Building an AdS/CFT superconductor}, Phys. Rev. Lett. \textbf{101} (2008)
031601 [arXiv:0803.3295].

\bibitem{Har2} S. A. Hartnoll, C. P. Herzog and G. T. Horowitz, \textit{%
Holographic superconductors}, JHEP \textbf{12} (2008) 015 [arXiv:0810.1563].

\bibitem{Hor} G. T. Horowitz, \textit{Introduction to holographic
superconductors}, Lect. Notes Phys. \textbf{828} (2011) 313
[arXiv:1002.1722].

\bibitem{Mus} D. Musso, \textit{Introductory notes on holographic
superconductors}, arXiv:1401.1504.

\bibitem{Gr} R. Gregory, S. Kanno and J. Soda, \textit{Holographic
superconductors with higher curvature corrections}, JHEP \textbf{0910}
(2009) 010 [arXiv:0907.3203].

\bibitem{RGC1} R. G. Cai, L. L, Li-Fang Li, R. Yang, \textit{Introduction to
holographic superconductor models}, Sci. China Phys. Mech. Astron. \textbf{58%
} (2015) 060401 [arXiv:1502.00437].

\bibitem{P.BGRL} R. Banerjee, S. Gangopadhyay, D. Roychowdhury, and A. Lala, 
\textit{Holographic s-wave condensate with nonlinear electrodynamics: A
nontrivial boundary value problem}, Phys. Rev. D \textbf{87} (2013) 104001
[arXiv:1208.5902].

\bibitem{P.MRM} D. Momeni, M. Raza, and R. Myrzakulov, \textit{More on
Superconductors via Gauge/Gravity Duality with Nonlinear Maxwell Field},
Journal of Gravity \textbf{2013} (2013) Article ID 782512 [arXive:1305.3541].

\bibitem{P.CW} C. M. Chen and M. F. Wu, \textit{An analytic analysis of
phase transitions in holographic superconductors}, Prog. Theor. Phys. 
\textbf{126} (2011) 387 [arXiv:1103.5130].

\bibitem{P.ZGJZ} H. B. Zeng, X. Gao, Y. Jiang, and H. S. Zong, \textit{%
Analytical computation of critical exponents in several holographic
superconductors}, JHEP \textbf{1105} (2011) 002 [arXiv:1012.5564].

\bibitem{RGC2} R. G. Cai, Z.Y. Nie, H.Q. Zhang, \textit{Holographic p-wave
superconductors from Gauss-Bonnet gravity}, Phys. Rev. D \textbf{82} (2010)
066007 [arXiv:1007.3321].

\bibitem{RGC3} R. G. Cai, H. F. Li, H. Q. Zhang, \textit{Analytical studies
on holographic insulator/superconductor phase transitions}, Phys. Rev. D 
\textbf{83} (2011) 126007 [arXiv:1103.5568].

\bibitem{RGC4} R. G. Cai, L. Li, L. F. Li, \textit{A holographic p-wave
superconductor model}, JHEP \textbf{1401} (2014) 032 [arxiv:1309.4877].

\bibitem{XHWang} X. H. Ge, S. F. Tu, B. Wang, \textit{d-Wave holographic
superconductors with backreaction in external magnetic fields}, JHEP \textbf{%
09} (2012) 088 [arXiv:1209.4272].

\bibitem{Wang2} X. M. Kuang, E. Papantonopoulos, G. Siopsis, B. Wang, 
\textit{Building a Holographic Superconductor with Higher-derivative
Couplings}, Phys. Rev. D \textbf{88} (2013) 086008 [arXiv:1303.2575].

\bibitem{Wang3} Q. Y. Pan and B. Wang, \textit{General holographic
superconductor models with Gauss-Bonnet corrections}, Phys. Lett. B \textbf{%
693} (2010) 159 [arXiv:1005.4743].

\bibitem{Wang4} Q. Pan, J. Jing, B. Wang, S. Chen, \textit{Analytical study
on holographic superconductors with backreactions}, JHEP \textbf{06} (2012)
087 [arXiv:1205.3543].

\bibitem{Wang5} Q. Pan, J. Jing, B. Wang, \textit{Analytical investigation
of the phase transition between holographic insulator and superconductor in
Gauss-Bonnet gravity}, JHEP \textbf{11} (2011) 088 [arXiv:1105.6153].

\bibitem{Wang6} Q. Pan, B. Wang, E. Papantonopoulos, J. Oliveira, A. B.
Pavan, \textit{Holographic Superconductors with various condensates in
Einstein-Gauss-Bonnet gravity}, Phys. Rev. D \textbf{81} (2010) 106007
[arXiv:0912.2475].

\bibitem{Wang7} M. Kord Zangeneh, Y. C. Ong, B. Wang, \textit{Entanglement
Entropy and Complexity for One-Dimensional Holographic Superconductors},
Phys. Lett. B \textbf{771} (2017) 235 [arXiv:1704.00557].

\bibitem{JJ1} J. Jing, Q. Pan, S. Chen, \textit{Holographic
Superconductor/Insulator Transition with logarithmic electromagnetic field
in Gauss-Bonnet gravity}, Phys. Lett. B \textbf{716} (2012) 385 [
arXiv:1209.0893].

\bibitem{JJ3} S. Chen, Q. Pan, J. Jing, \textit{Holographic superconductors
in quintessence AdS black hole,} Class. Quantum Grav. \textbf{30} (2013)
145001 [arXiv:1206.2069].

\bibitem{SG2} S. Gangopadhyay and D. Roychowdhury, \textit{Analytic study of
properties of holographic p-wave superconductors}, JHEP \textbf{08} (2012)
104 [arXiv:1207.5605].

\bibitem{Shey2} A. Sheykhi, H. R. Salahi, A. Montakhab, \textit{Analytical
and Numerical Study of Gauss-Bonnet Holographic Superconductors with
Power-Maxwell Field}, JHEP \textbf{04} (2016) 058 [arXiv:1603.00075].

\bibitem{Shey3} H. R. Salahi, A. Sheykhi, A. Montakhab, \textit{Effects of
Backreaction on Power-Maxwell Holographic Superconductors in Gauss-Bonnet
Gravity} Eur. Phys. J. C \textbf{76} (2016) 575 [arXiv:1608.05025].

\bibitem{Shey4} A. Sheykhi, F. Shaker, \textit{Analytical study of
properties of holographic superconductors with exponential nonlinear
electrodynamics}, Canadian J. of Phys. \textbf{94} (12) (2016) 1372
[arXiv:1601.05817].

\bibitem{SheyLN} A. Sheykhi, F. Shamsi, \textit{Holographic Superconductors
with Logarithmic Nonlinear Electrodynamics in an External Magnetic Field},
Int. J. Theor. Phys. \textbf{56} (2017) 916 [arXiv:1603.02678].

\bibitem{SheyPM} A. Sheykhi, F. Shamsi, S. Davatolhagh, \textit{The upper
critical magnetic field of holographic superconductor with conformally
invariant Power--Maxwell electrodynamics}, Can. J. Phys. \textbf{95} (2017)
450 [arXiv:1609.05040].

\bibitem{1609.08402} B. Pourhassan and M. M. Bagheri-Mohagheghi, \textit{%
Holographic superconductor in a deformed four-dimensional STU model},
arXiv:1609.08402.

\bibitem{JJ2} J. Jing, S. Chen, \textit{Holographic superconductors in the
Born-Infeld electrodynamics,} Phys. Lett. B \textbf{686} (2010) 68
[arXiv:1001.4227].

\bibitem{Gan1} S. Gangopadhyay, D. Roychowdhury, \textit{\ Analytic study of
properties of holographic superconductors in Born-Infeld electrodynamics},
JHEP \textbf{05} (2012) 002 [arXiv:1201.6520].

\bibitem{Gan2} S. Gangopadhyay, D. Roychowdhury, \textit{\ Analytic study of
Gauss-Bonnet holographic superconductors in Born-Infeld electrodynamics}
JHEP \textbf{05} (2012) 156 [arXiv:1204.0673].

\bibitem{Zh} Z. Zhao, Q. Pan, S. Chen and J. Jing, \textit{Notes on
holographic superconductor models with the nonlinear electrodynamics}, Nucl.
Phys. B \textbf{871} (2013) 98 [arXiv:1212.6693].

\bibitem{1306.0064} W. Yao and J. Jing, \textit{Analytical study on
holographic superconductors for Born-Infeld electrodynamics in Gauss-Bonnet
gravity with backreactions}, JHEP \textbf{1305} (2013) 101 [arXiv:1306.0064].

\bibitem{Lai} C. Lai, Q. Pan, J. Jing, Y. Wang, \textit{On analytical study
of holographic superconductors with Born-Infeld electrodynamics}, Phys.
Lett. B \textbf{749} (2015) 437 [arXiv:1508.05926].

\bibitem{SG1} D. Ghorai, S. Gangopadhyay, \textit{Higher dimensional
holographic superconductors in Born-Infeld electrodynamics with backreaction}%
, Eur. Phys. J. C \textbf{76} (2016) 146 [arXiv:1511.02444]

\bibitem{Shey1} A. Sheykhi, F. Shaker, \textit{Analytical study of
holographic superconductor in Born-Infeld electrodynamics with backreaction}%
, Phys. Lett. B \textbf{754} (2016) 281 [arXiv:1601.04035].

\bibitem{1211.0904} D. Roychowdhury, \textit{Effect of external magnetic
field on holographic superconductors in presence of nonlinear corrections},
Phys. Rev. D \textbf{86} (2012) 106009 [arXiv:1211.0904].

\bibitem{1710.09630} D. Ghorai and S. Gangopadhyay, \textit{Conductivity of
holographic superconductors in Born-Infeld electrodynamics},
arXiv:1710.09630.

\bibitem{Lif} S. Kachru, X. Liu and M. Mulligan, \textit{Gravity Duals of
Lifshitz-like Fixed Points}, Phys. Rev. D \textbf{78} (2008) 106005
[arXiv:0808.1725].

\bibitem{masssol} G. Bertoldi, B. A. Burrington and A. Peet, \textit{Black
holes in asymptotically Lifshitz spacetimes with arbitrary critical exponent}%
, Phys. Rev. D \textbf{80} (2009) 126003 [arXiv:0905.3183];\newline
M. H. Dehghani and R. B. Mann, \textit{Lovelock-Lifshitz Black Holes}, JHEP 
\textbf{1007} (2010) 019 [arXiv:1004.4397].

\bibitem{Deh1} M. H. Dehghani and R. B. Mann, \textit{Thermodynamics of
Lovelock-Lifshitz black branes}, Phys. Rev. D \textbf{82 }(2010) 064019
[arXiv:1006.3510];\newline
M. H. Dehghani and Sh. Asnafi, \textit{Thermodynamics of rotating
Lovelock-Lifshitz black branes}, Phys. Rev. D \textbf{84 }(2011) 064038
[arXiv:1107.3354];\newline
M. H. Dehghani, Ch. Shakuri and M. H. Vahidinia, \textit{Lifshitz black
brane thermodynamics in the presence of a nonlinear electromagnetic field},
Phys. Rev. D \textbf{87 }(2013) 084013 [arXiv:1306.4501].

\bibitem{hcc} M. Bravo-Gaete and M. Hassaine, \textit{Thermodynamics of
charged Lifshitz black holes with quadratic corrections}, Phys. Rev. D 
\textbf{91 }(2015) 064038 [arXiv:1501.03348].

\bibitem{toplif} R. B. Mann, \textit{Lifshitz Topological Black Holes}, JHEP 
\textbf{0906} (2009) 075 [arXiv:0905.1136].

\bibitem{peet} G. Bertoldi, B. A. Burrington and A. W. Peet, \textit{Thermal
behavior of charged dilatonic black branes in AdS and UV completions of
Lifshitz-like geometries}, Phys. Rev. D \textbf{82 }(2010) 106013
[arXiv:1007.1464].

\bibitem{tario} J. Tarrio and S. Vandoren, \textit{Black holes and black
branes in Lifshitz spacetimes}, JHEP \textbf{1109 }(2011) 017
[arXiv:1105.6335].

\bibitem{Deh2} M. Kord Zangeneh, A. Sheykhi and M. H. Dehghani, \textit{%
Thermodynamics of topological nonlinear charged Lifshitz black holes}, Phys.
Rev. D \textbf{92} (2015) 024050 [arXiv:1506.01784].

\bibitem{Deh3} M. Kord Zangeneh, M. H. Dehghani and A. Sheykhi, \textit{%
Thermodynamics of Gauss-Bonnet-Dilaton Lifshitz Black Branes}, Phys. Rev. D 
\textbf{92} (2015) 064023 [arXiv:1506.07068].

\bibitem{hololif} M. Taylor, \textit{Non-relativistic holography},
arXiv:0812.0530;\newline
D. Momeni, R. Myrzakulov, L. Sebastiani and M. R. Setare, \textit{Analytical
holographic superconductors in }$AdS_{N}$\textit{-Lifshitz topological black
holes}, Int. J. Geom. Meth. Mod. Phys. \textbf{12} (2015) 1550015
[arXiv:1210.7965];\newline
D. Roychowdhury, \textit{Lifshitz holography and the phases of the
anisotropic plasma}, arXiv:1509.05229.

\bibitem{vislif} D. W. Pang, \textit{On Charged Lifshitz Black Holes}, JHEP 
\textbf{1001} (2010) 116 [arXiv:0911.2777];\newline
A. Bhattacharyya and D. Roychowdhury, \textit{Lifshitz Hydrodynamics And New
Massive Gravity}, arXiv:1503.03254.

\bibitem{CondLif} S. A. Hartnoll, D. M. Hofman and D. Vegh, \textit{Stellar
spectroscopy: Fermions and holographic Lifshitz criticality}, JHEP \textbf{%
1108} (2011) 096 [arXiv:1105.3197].

\bibitem{KordLif1} M. Kord Zangeneh, A. Dehyadegari, A. Sheykhi and M.H.
Dehghani, \textit{Thermodynamics and gauge/gravity duality for Lifshitz
black holes in the presence of exponential electrodynamics}, JHEP \textbf{%
1603} (2016) 037 [arXiv:1601.04732]

\bibitem{KordLif2} A. Dehyadegari, A. Sheykhi and M. Kord Zangeneh, \textit{%
Holographic Conductivity for Logarithmic Charged Dilaton-Lifshitz Solutions}%
, Phys. Lett. B \textbf{758} (2016) 226 [arXiv:1602.08476].

\bibitem{KordLif3} M. Kord Zangeneh, A. Dehyadegari, M. R. Mehdizadeh, B.
Wang, A. Sheykhi, \textit{Thermodynamics, phase transitions and Ruppeiner
geometry for Einstein-dilaton Lifshitz black holes in the presence of
Maxwell and Born-Infeld electrodynamics}, Eur. Phys. J. C \textbf{77} (2017)
423 [arXiv:1610.06352].

\bibitem{KordLif4} M. Kord Zangeneh, B. Wang, A. Sheykhi, Z. Y. Tang, 
\textit{Charged scalar quasi-normal modes for linearly charged
dilaton-Lifshitz solutions}, Phys. Lett. B \textbf{771} (2017) 257
[arXiv:1701.03644].

\bibitem{Bry} E.J. Brynjolfsson, U.H. Danielsson, L. Thorlacius, T. Zingg, 
\textit{Holographic Superconductors with Lifshitz Scaling} J. Phys. A 
\textbf{43} (2010) 065401 [arXiv:0908.2611].

\bibitem{Sin} S. J. Sin, S. S. Xu and Y. Zhou, \textit{Holographic
Superconductor for a Lifshitz fixed point}, Int. J. Mod. Phys. A \textbf{26}
(2011) 4617 [arXiv:0909.4857].

\bibitem{Bu} Y. Bu, \textit{Holographic superconductors with z=2 Lifshitz
scaling}, Phys. Rev. D \textbf{86} (2012) 046007 [arXiv:1211.0037].

\bibitem{Lu} J. W. Lu, Y. B. Wu, P. Qian, Y. Y. Zhao, X. Zhang and N. Zhang, 
\textit{Lifshitz Scaling Effects on Holographic Superconductors}, Nucl.
Phys. B \textbf{887 }(2014)112 [arXiv:1311.2699].

\bibitem{Tall} G. Tallarita, \textit{Holographic Lifshitz Superconductors
with an Axion Field}, Phys. Rev. D \textbf{89} (2014) 106005
[arXiv:1402.4691].

\bibitem{Mirza1} S. A. Hosseini Mansoori, B. Mirza, A. Mokhtari, F.
Lalehgani Dezaki, Z. Sherkatghanad, \textit{Weyl holographic superconductor
in the Lifshitz black hole background}, JHEP \textbf{1607} (2016) 111
[arXiv:1602.07245].

\bibitem{Mirza2} Z. Sherkatghanad, B. Mirza, F. Lalehgani Dezaki, \textit{%
Exponential nonlinear electrodynamics and backreaction effects on
Holographic superconductor in the Lifshitz black hole background}, Int. J.
of Mod. Phys. D \textbf{26} (2017) 1750175 [arXiv:1708.04289].

\bibitem{1008.0720} X. Gao and H.-B. Zhang, \textit{Refractive index in
holographic superconductors}, JHEP \textbf{1008} (2010) 075
[arXiv:1008.0720].

\bibitem{1107.1242} A. Amariti, D. Forcella, A. Mariotti and M. Siani, 
\textit{Negative Refraction and Superconductivity}, JHEP \textbf{1110}
(2011) 104 [arXiv:1107.1242].

\bibitem{1305.6273} S. Mahapatra, P. Phukon and T. Sarkar, \textit{%
Generalized Superconductors and Holographic Optics}, JHEP \textbf{1401}
(2014) 135 [arXiv:1305.6273].

\bibitem{1404.2190} A. Dey, S. Mahapatra and T. Sarkar, \textit{Generalized
Holographic Superconductors with Higher Derivative Couplings}, JHEP \textbf{%
1406 }(2014) 147 [arXiv:1404.2190].

\bibitem{1411.6405} S. Mahapatra, \textit{Generalized Superconductors and
Holographic Optics - II}, JHEP \textbf{1501 }(2015) 148 [arXiv:1411.6405].

\bibitem{0905.2678} D.-W. Pang, \textit{A Note on Black Holes in
Asymptotically Lifshitz Spacetime}, arXiv:0905.2678.

\bibitem{DL} R. A. Depine and A. Lakhtakia, \textit{A new condition to
identify isotropic dielectric-magnetic materials displaying negative phase
velocity}, Micro. Opt. Tech. Lett. \textbf{41} (2004) 315.

\bibitem{horo} G. T. Horowitz and M. M. Roberts, \textit{Holographic
Superconductors with Various Condensates}, Phys. Rev. D \textbf{78} (2008)
126008 [arXiv:0810.1077].

\bibitem{landau} L. D. Landau and E. M. Lifshitz, \textit{Electrodynamics of
Continous Media}, Pergamon Press, Oxford (1984).

\bibitem{son} D. T. Son and A. O. Starinets, \textit{Minkowski space
correlators in AdS/CFT correspondence: Recipe and applications}, JHEP 
\textbf{0209} (2002) 042 [hep-th/0205051].

\bibitem{mark} V. A. Markel, \textit{Can the imaginary part of permeability
be negative ?}, Phys. Rev. E \textbf{78} (2008) 026608.

\bibitem{zhao} Q. Zhao, L. Kang, B. Li and J. Zhou, \textit{Tunable negative
refraction in nematic liquid crystals}, App. Phys. Lett. \textbf{89} (2006)
221918.
\end{thebibliography}
\end{document}